\documentclass[12pt]{article}
\usepackage{tikz}
\usepackage{amsmath}
\usepackage{mathtools}
\usepackage{fancyhdr}
\usepackage{gensymb}
\usepackage{graphicx}
\usepackage{cite}
\usetikzlibrary{decorations.pathmorphing, arrows, patterns}
\tikzset{snake it/.style={-stealth,
decoration={snake, 
    amplitude = .4mm,
    segment length = 2mm,
    post length=0.9mm},decorate}}
\usepackage{amssymb}

\usetikzlibrary{decorations.markings}

\usepackage[a4paper, total={6in, 10in}]{geometry}

\begin{document}
\title{\Large{\textbf{On the scattering Aharonov-Bohm effect}}}
\date{}
\author{Boris Iveti\'c
\footnote{bivetic@yahoo.com}
 \\ Department of Physics, Faculty of Science, University of Zagreb
  \\ Bijeni\v cka cesta 32, 10000 Zagreb, Croatia \\}

\maketitle

\begin{abstract}
In this paper we review some aspects of the scattering Aharonov-Bohm effect and Berry's phase. Specifically,  the problem of scattering of free 2d electrons on the system of an arbitrary number of parallel, infinitely thin and infinitely long coils (non-interacting, point magnetic fluxes) is modeled as free semi-classical particle propagation on a topologically nontrivial background. We show that in this case it is possible to obtain Berry's phase by integration. First the case of a single coil is analyzed, upon which a particular solution for the case of scattering on an arbitrary number of coils is found. Considering the solution in the momentum representation,  an analytic continuation of the angle coordinate in the momentum space is introduced, in accord with the nontrivial geometry of the configuration space. Finally, some simple experiments with possible applications for quantum computing are proposed.   
\end{abstract}

\section{Introduction}
In their seminal paper, Aharonov and Bohm \cite{AB} solved the problem of scattering of free electrons on infinitely thin and infinitely long coil carrying constant current, producing a field-free vector potential outside the coil. The solution wave-function in coordinate space in terms of Bessel functions of generally non-integer (and non-half-integer) order suffers from non-single-valuedness at the branch line.    

There exist theoretical explanations for this effect, most notably in terms of  Berry's phase (holonomies) \cite{ber,sim}, anyonic nature of such systems \cite{wilczek}, as well as through gauge and other symmetries \cite{jack}. Notably, the physical meaning of the phase of a wave-function was discussed already by Born and Fock \cite{BF} and  Dirac \cite{dirac} at the very beginnings of the (new) quantum theory. It is also interesting that meanwhile the analyses of geometrical phases found application in a variety of very different physical contexts, from anomalies in field theory to Hall's effect in solid state physics 
\cite{jack,revija, knjiga}, with high expectations for application in future quantum computing \cite{qc}.

However, the relations between different approaches, and different contexts in which the effect appears are not completely understood (see review in \cite{revija}), the behavior of phase for non-cyclic evolutions is barely discussed, and the case of $n\ge2$ parallel coils is not yet solved \cite{nambu, sto}. We also note that the analysis of the problem in the momentum representation, apart from few exceptions \cite{nambu,ganga,drago} is almost absent as opposed to coordinate representation.\\

Here we study the interaction of electrons with field-free vector potential as a propagation of free particles on the topologically nontrivial background configuration space, for which cotangent spaces at different points can not be identified. We show that motion on this background produces a phase that is single-valued and integrable, the familiar discontinuity emerging upon the projection of the solution on real Euclidean space. Furthermore, we show that instead of variation of an external parameter of the Hamiltonian, one can also arrive at Berry's phase by considering variation of an internal parameter, that is the cotangent space of the configurational space, as the particles propagate over it (compare to \cite{AA}).

In section 2 we lay down the procedure for obtaining Berry's phase on example of plane waves scattering on a single coil, reproducing also the general solution of Aharonov and Bohm \cite{AB}, and analyzing the case of multiple coils. Section 3 studies the solution in the momentum representation, showing the phase does not depend on the representation. In section 4 some simple experimental proposals to test our findings, as well as their possible practical applications in quantum computing, are shortly discussed.

\section{Configuration space}
Consider the action of two infinitesimal displacements operators $\nabla$ and $\nabla_c$, with eigenvalues $\mathbf P$ and $\mathbf p$,

\begin{equation}\label{flat space displacement}
\nabla\psi(\mathbf{r;P})=\mathbf P \psi(\mathbf{r;P})
\end{equation}

\begin{equation}\label{curved space disp}
\nabla_c\psi(\mathbf{r;p})=\mathbf p \psi(\mathbf{r;p}).
\end{equation}

In particular, consider the case where the spectra of the squares of the two operators are the same,

\begin{equation}\label{Sch.flat}
-\nabla^2 \psi(\mathbf{r; P})=2mE\psi(\mathbf{r; P})
\end{equation}

\begin{equation}\label{Sch. curved}
-\nabla_c^2 \psi(\mathbf{r; p})=2mE\psi(\mathbf{r; p}),
\end{equation}



It is obvious that the operators $\nabla$ and $\nabla_c$ can not act on the same vector space, otherwise their eigenvalues would be identical. We take operator $\nabla$ to be the standard infinitesimal displacement operator acting on vectors on Euclidean space, $\psi(\mathbf{r;P})\in L^2(\mathbb{R}^2)$. The operator $\nabla_c$ is a \textit{covariant} momentum, a global infinitesimal displacement operator acting on vectors defined on non-trivial background ($\psi(\mathbf{r; p})\in L^2(\mathbb{R_+\times R})$ \cite{sto} for a single coil), and belonging to the cotangent space at the point where the wave-function identically vanishes. On this background, operator $\nabla$ is defined as the element of the position dependent cotangent space, and is given as a combination of displacement and multiplication operators, 

\begin{equation}
\nabla(\mathbf{r})=\nabla_c-ie\mathbf{A(r)},
\end{equation}
where $\mathbf{A(r)}$ is the introduced vector potential whose rotation vanishes, in this paper considered to be 

\begin{equation}
e\mathbf{A(r)}=\sum_i\frac{\alpha_i}{|\mathbf{r-a_i}|} \frac{\mathbf{\hat z\times (r-a_i)}}{|\mathbf{r-a_i|}},
\end{equation}
with $\alpha_i$ real constants equal to the strengths of point magnetic fluxes at positions $\mathbf{r=a_i} $. It is the displacement operator for vectors defined on flat space constructed around some point $\mathbf r$, or on the projection of non-trivial background onto Euclidean space, i.e.

\begin{equation}\label{projection}
(-i\nabla+e\mathbf{A(r)})^2\psi^{pr}(\mathbf{p;r})=2mE\psi^{pr}(\mathbf{p;r})
\end{equation}
where $\psi^{pr}(\mathbf{r; p})\in L^2(\mathbb{R}^2\backslash\lbrace\text{cuts} \rbrace)$.

Non-standard geometry of the configuration space is also seen in the algebra of the operators. While the Heisenberg algebra for the operators $\nabla$ and $\mathbf r$ is a standard one, i.e.

\begin{equation}
[r_i,r_j]=0, \ \ \ [r_i,P_j]=i\delta_{ij}, \ \ \ [P_i,P_j]=0,
\end{equation}
the algebra of operators $\nabla_c$ and $\mathbf r$ gets deformed,

\begin{equation}
[r_i,r_j]=0, \ \ \ [r_i,p_j]=i\delta_{ij}, \ \ \ [p_i,p_j]\neq 0.
\end{equation}
Noncommutativity of infinitesimal displacements in different directions locally corresponds to a global effect of the non-vanishing phase for cyclic evolutions.

Some elements of the geometry of $\psi(\mathbf{r; p})\in L^2(\mathbb{R_+\times R})$ are given in \cite{sto}. Most importantly, the metric is unchanged, meaning that the phase can be simply integrated

\begin{equation}\label{connection}
\psi(\mathbf{r;p})=\psi(\mathbf{r;P})|_{\mathbf{P=p}+\mathbf{A(r)}}\exp\left(i\int_{\mathbf{P=p}(\mathbf{r'=\infty})}^{\mathbf{P=p}+\mathbf{A(r)}(\mathbf{r'=r})} (\mathbf{p-P})d\mathbf{r'}\right).
\end{equation} 
This follows from the fact that the spectra of the squares of the two operators are the same, meaning their eigenvectors, when projected onto each other, can differ only by an unit phase. It also requires the existence of a part of the domain where the two sets of eigenvectors are identical. This is achieved by extending the domain with a so called ideal point \cite{sto}, where $\mathbf{P=p}$, which we take to be $(r=\infty,\varphi= \varphi_0)$ but also discuss other possibilities.

 One can also arrive at (\ref{connection}) by simply introducing Ansatz $\psi^{pr}(\mathbf{r;p})=\psi(\mathbf{r;P})|_{\mathbf{P=p}+\mathbf{A(r)}}e^{iG(\mathbf r)}$ into (\ref{projection}), and evaluate the unknown function $G(\mathbf{r})$ to be as given in (\ref{connection}) upon inverse projection (extension).

\subsection{Single coil} 
In this case

\begin{equation}
\mathbf{A(r)}=\frac{\alpha}{r}\hat\varphi,
\end{equation}
with magnetic flux of strength $\alpha$ at the origin.

A particular solution of (\ref{Sch.flat}) that will be considered in practical application is a simple plane wave, $\psi(\mathbf{r;P})=e^{i\mathbf{Pr}}$, so we start with it. From (\ref{connection}) is seen that the radial component of the difference of the displacements vanishes, so that the radial motion doesn't acquire a phase. Only the angular motion matters, giving

\begin{equation}\label{plane wave}
\psi(\mathbf{r;p})=e^{i\mathbf{pr}}e^{-i\alpha(\varphi-\varphi_0)},
\end{equation}
which is a particular solution of (\ref{Sch. curved}). The phase factor is Berry's phase, an angle between vectors as they are parallel transported from point at infinity (point where interaction vanishes) to some point $\mathbf{r}$, via operators $\nabla$ and $\nabla_c$. The solution is single valued, since it is defined on the domain $L^2(\mathbb{R_+\times R})$.

Irrelevance of the radial motion for the phase means that the ideal point could be taken anywhere on the line $\varphi=\varphi_0$. In particular one may consider defining ideal point by extending the domain with the point $(r=0, \varphi=\varphi_0)$ \cite{sto}. This allows for an interpretation of the phase as coming from parallel transport of cotangent space of a point flux to the position of an electron. Note that our considerations are directly applicable to localized systems, whose background domain is an open subset of $\mathbb{R}^2$. In this case ideal point is taken to be on the boundary of the domain, and is identified with Berry's external parameter $\mathbf R$ \cite{ber}.

Of some interest might be to reproduce the full solution of Aharonov and Bohm \cite{AB}, in terms of standing waves (Bessel functions) that vanish at infinity. Remembering that the usual derivation of Bessel's standing waves comes from separation of the variables, $\psi(\mathbf{r;P})=R(r;\mathbf{P})\Phi(\varphi;\mathbf{P})$, after which 

\begin{equation}\label{angular free}
-\frac{\partial^2}{\partial \varphi^2}\Phi(\varphi;\mathbf{P})=M^2\Phi(\varphi;\mathbf{P})
\end{equation}

\begin{equation}\label{radial free}
\left( r^2\frac{\partial^2}{\partial r^2}+r\frac{\partial}{\partial r}+2mEr^2\right)R(r;\mathbf{P})=M^2R(r;\mathbf{P})
\end{equation}
The single-valuedness of the angular part that fixes $M$ to be a non-negative integer, upon which follows the general solution

\begin{equation}\label{gen sol free}
\psi(\mathbf{r;P})=\sum_M a_M J_M(\sqrt{2mE}r)e^{iM\varphi}. 
\end{equation}

When vector potential is introduced, equation (\ref{angular free}) changes

\begin{equation}\label{angular interaction}
-\left(\frac{\partial}{\partial \varphi}+i\alpha\right)^2\Phi(\varphi;\mathbf{p})=M^2\Phi(\varphi;\mathbf{p}),
\end{equation}

implying that the change $\mathbf{P\to p}+\frac{\alpha}{r}\hat\varphi$ corresponds to changing $M\to |m+\alpha|$. This can also be seen from the change in the angular momentum operator,

\begin{equation}\label{ang mom}
\mathbf{J=r\times P}=\frac{\partial}{\partial\Phi}\hat z\to \mathbf{r\times p}+\alpha\hat z=\left(\frac{\partial}{\partial\varphi}+\alpha\right)\hat z=\mathbf{j}+\alpha\hat z,
\end{equation}
 where operators $\mathbf{J}$ and $\mathbf{j}$ have positive integers $M$ and $m$ as eigenvalues and act on vectors on the background of $\mathbb{R}^2$ and $\mathbb{R_+\times R}$ respectively. 

From (\ref{connection}) then follows the general solution of Aharonov and Bohm:

\begin{align}\label{full interaction}
\psi(\mathbf{r;p})&=\psi(\mathbf{r;P})|_{M=|m+\alpha|}\exp
\left(i\int_{M=m(\varphi'=\varphi_0)}^{M=m+\alpha(\varphi'=\varphi)} 
(m-M)d\varphi'\right)\\
&=e^{i\alpha\varphi_0}\sum_m a_m J_{|m+\alpha|}(\sqrt{2mE}r)e^{im\varphi}.
\end{align}
Projecting it on $\mathbb{R}^2$, $\psi^{pr}(\mathbf{r;p})$ has a branch line (cut) at $\varphi=\varphi_0$.

\subsection{Multiple coils}
The rotational symmetry that was deformed by the introduction of a single coil disappears when other coils are added, as in this case angular and radial part of the equation don't separate. This makes the analyses of the full solution difficult. But a particular solution of plane waves, which are not eigenstates of the angular momentum operator, is directly obtainable from (\ref{connection}),

\begin{equation}
\psi(\mathbf{p;r})=e^{i(\mathbf{p+A(r)})\mathbf r} \prod_{i=1}^ne^{-i\alpha_i(\varphi-\varphi_0)_i}.
\end{equation}
Here $(\varphi-\varphi_0)_i$ is an angle that the ray connecting $i$-th point flux and $\mathbf r$ makes with the a ray connecting $i$-th point flux and the ideal point. When projected onto $\mathbb{R}^2$ background, the solution $\psi^{pr}(\mathbf{r;p})$ has $n$ cuts, going from positions of the $n$ fluxes to the ideal point at infinity $(r=\infty, \varphi=\varphi_0)$. When considering a localized system, the phase it gets in a cyclic evolution is equal to the total encircled flux times $2\pi$.

In this case the background geometry is more complicated then in the case of a single coil. Separability of phases due to different point fluxes,

\begin{equation}
\psi(\mathbf{p;r})=e^{i\mathbf{p}\mathbf r} \prod_{i=1}^ne^{\mathbf{A_i(r)a_i}}e^{-i\alpha_i(\varphi-\varphi_0)_i},
\end{equation}
where $\mathbf{A_i(r)}$ is the strength of the vector potential at point $\mathbf r$ from $i$-th point flux only, allows for a highly formal expression for the domain,

\begin{equation}
\psi(\mathbf{r;p})=\in L^2\left(\bigcup_{i=1}^n (\mathbb{R_+\times R})_i\right),
\end{equation} 
where $(\mathbb{R_+\times R})_i$ is $\mathbb{R_+\times R}$ space around the $i$-th flux, and with the understanding that the full phase is a sum of individual phases.

\section{Momentum representation}
Identifying Berry's phase with the angle that a vector from cotangent space makes as it is parallel transported from the point at infinity (point of no interaction) to some point should be independent of the representation.  We proceed to show that on an example of plane waves.
 

In the absence of flux we have
\begin{equation}
\mathbf{P^2}\tilde\Psi(\mathbf P)=2mE\tilde\Psi(\mathbf P),
\end{equation}
with the particular solution
\begin{equation}
\tilde\Psi(\mathbf P)=\frac{1}{P} \delta(P-\sqrt{2mE})\delta(\Theta-\Theta_0),
\end{equation}
where $\Theta_0$ is the angle incoming electrons make with $x$-axes. When flux is introduced, equation becomes

\begin{equation}
\left(\mathbf p + q\int d\mathbf p' \mathbf{A(p-p')}    \right)^2\tilde\Psi(\mathbf p)=2mE\tilde\Psi(\mathbf p),
\end{equation}

where $\mathbf{A(p-p')}$ is the Fourier transform of the vector potential,

\begin{equation}
q\mathbf{A(p)}=\frac{\alpha}{p}\hat\theta.
\end{equation}

In coordinate representation differential operator was replaced by a differential operator plus multiplication; in the momentum representation, multiplication operator is replaced by multiplication plus integral operator,

\begin{equation}\label{repl mom}
\mathbf{P=p}+\int d\mathbf p'\frac{\alpha}{|\mathbf{p-p'}|}\hat\varphi, 
\end{equation}
which makes the analyses more difficult. Note that the above highly formal notation hides the fact that integral operator is not local (it acts on $\tilde\Psi(\mathbf{p'})$), and also that the square of the vector potential vanishes upon Fourier transformation. Following (\ref{connection}), we write

\begin{equation}
\tilde\Psi(\mathbf p)=\tilde\Psi(\mathbf P)|_{\mathbf{P=p}+\int d\mathbf p'\frac{\alpha}{|\mathbf{p-p'}|}\hat\varphi}\exp\left(i\int_{\mathbf{P=p}(\mathbf{r'=\infty})}^{\mathbf{P=p}+\frac{\alpha}{r}\hat\varphi(\mathbf{r'=r})} (\mathbf{P-p})d\mathbf{l'}\right).
\end{equation}

It is particularly difficult to implement the condition (\ref{repl mom}). The radial component (square) of the momentum doesn't change, so that $P$ is everywhere to be replaced simply by $p$. The angular component doesn't change either; we argue heuristically that the only change is the extension of the domain of the angular component of the momentum, due to nontrivial geometry of the configurational space. 

That is

\begin{equation}\label{mom sol}
\tilde\Psi(\mathbf p)=\frac{1}{p} \delta(p-\sqrt{2mE})\delta(\theta-\theta_0)e^{-i\alpha(\varphi-\varphi_0)}.
\end{equation}
and we proceed to treat cases of integer/non-integer $\alpha$ separately.
\begin{figure}[h]
\begin{center}
\begin{tikzpicture}[line cap=round,line join=round,>=triangle 45,x=1.0cm,y=1.0cm,scale=0.8]
\draw [->] (3.08,1.84) -- (4.38,1.84);
\draw [->] (-3.32,0.54) -- (2,0.54);
\draw [->] (5.58,0.54) -- (12.34,0.54);
\draw [->,line width=2pt] (7.5,0.54) -- (7.5,4.6);
\draw [->,line width=2pt] (9.9,0.54) -- (9.9,4.6);
\draw[line width=2pt] (7.5,0.54)--(9.9,0.54);
\draw (-2.1,-0.2) node {$ -\pi $};
\draw (0.7,-0.2) node {$ \pi $};
\draw (7.2,-0.2) node {$ -\pi $};
\draw (9.9,-0.2) node {$ \pi $};
\draw (6,5) node {$ -\pi+i\infty $};
\draw (11.05,5) node {$ \pi+i\infty $};
\draw [->] (8.7,-0.4) -- (8.7,5.8);
\draw (-0.7,-1) node {$ \theta \in \langle -\pi, \pi\rangle $};
\draw (8.78,-1) node {$  \theta \in \langle -\pi+i\infty, \pi +i\infty \rangle  $};
\draw (2.06,0.08) node[anchor=north west] {$ \theta $};
\draw (12.1,0.28) node[anchor=north west] {$ \theta_R $};
\draw (7.8,5.9) node[anchor=north west] {$ \theta_I $};
\draw (-2,0.72)-- (-2,0.38);
\draw (0.6,0.72)-- (0.6,0.38);
\draw [line width=2pt] (-2,0.54) -- (0.6,0.54);
\end{tikzpicture}
\end{center}
\caption{The standard momentum variable domain gets extended into the complex plane. The path of integration follows the thick line counter-clockwise.} \label{fig:M2}
\end{figure}
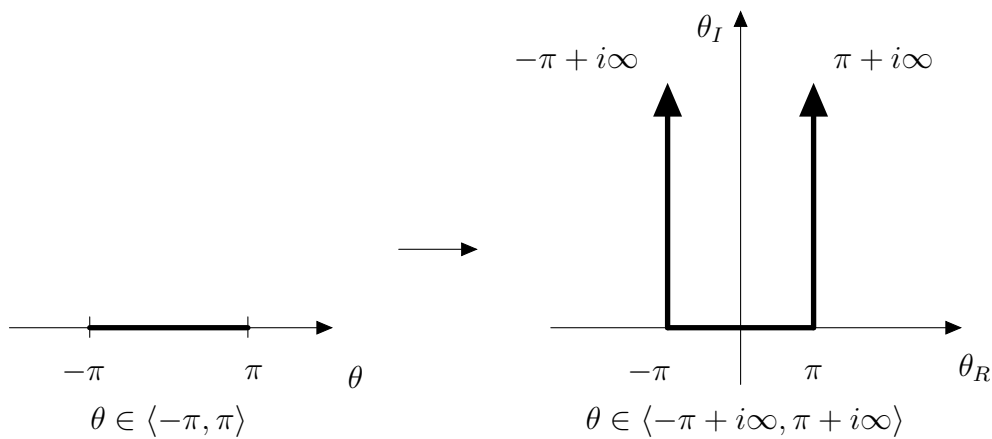

\subsection{Integer $\alpha$}
When $\alpha$ is an integer, the background geometry is equivalent to $\mathbb{R}^2$ and we use the standard 
inverse Fourier transform in two dimensions,

\begin{equation}\label{ft2d}
\mathcal{F}\lbrace\tilde\Psi(\mathbf{p})\rbrace=\frac{1}{2\pi}\int_{-\pi}^\pi d\theta\int_0^\infty pdp e^{-ipr\sin(\theta+\phi)}\tilde\Psi(\mathbf{p})\equiv \Psi(\mathbf r; \mathbf p),
\end{equation}
where $\phi\equiv \pi/2-\varphi$ is introduced for convenience.
Inserting the solution (\ref{mom sol}), doing the radial integral and representing the Dirac's delta in the angular part as a series,
\begin{equation}
\delta(x)=\frac{1}{2\pi}\sum_{m=-\infty}^\infty e^{imx},
\end{equation}
gives
\begin{align}\label{ft2d2}
\psi(\mathbf{r;p})&=\sum_{m=-\infty}^\infty(-i)^\alpha\frac{1}{2\pi}\int_{-\pi}^{\pi}e^{-ikr\sin(\theta+\phi)}e^{im(\theta-\theta_0)}e^{i\alpha(\theta+\phi-\theta_0+\varphi_0)}d\theta\\
&=\sum_{m=-\infty}^\infty(-i)^\alpha e^{-im\phi}e^{-i(m+\alpha)\theta_0}e^{i\alpha\varphi_0}\frac{1}{2\pi}\int_{-\pi+\phi}^{\pi+\phi}e^{-ikr\sin(\theta+\phi)}e^{i(m+\alpha)(\theta+\phi)}d(\theta+\phi)\\
&\equiv\sum_{m=-\infty}^\infty \psi_m(\mathbf{r;p}),
\end{align}
where in the first equality a factor $e^{i\alpha(\theta-\theta_0})$, equal to unity upon integration is inserted, and $k\equiv \sqrt{2mE}$ is introduced for short. Due to periodicity (single-valuedness) of the phase, this becomes

\begin{equation}\label{ft2d3}
\Psi(\mathbf{r;p})=\sum_{m=-\infty}^\infty(-i)^\alpha e^{-im\phi}\int_{-\pi}^{\pi}e^{-ikr\sin\theta}e^{i(m+\alpha)\theta}d\theta,
\end{equation}
and with a familiar integral representation of Bessel's function of non-negative integer order (e.g.\cite{GR}, pg 1399),

\begin{equation}\label{Bessel1}
J_n(z)=\frac{1}{2\pi}\int_{-\pi}^{\pi} e^{-iz\sin\theta+in\theta}d\theta,
\end{equation}
gives,

\begin{equation}\label{coorwf}
\Psi_{m}(\mathbf{r;p})=(-i)^\alpha ke^{-im\phi} J_{m+\alpha}(kr)
\end{equation}
for $m+\alpha$ positive, and

\begin{equation}\label{coorwfn}
\Psi_{m}(\mathbf{r;p})=(-i)^\alpha (-1)^{m+\alpha}ke^{-im\phi} J_{-m-\alpha}(kr)
\end{equation}
for $m+\alpha$ negative.\newline
Comparing with the plane-wave solution of Aharonov and Bohm \cite{AB},

\begin{equation}\label{wfAB}
\Psi_{AB}(\mathbf{r;p})= \sum_{m=-\infty}^{\infty} (-i)^{|m+\alpha|}J_{|m+\alpha|}(kr)e^{im\varphi},
\end{equation} 
one sees that it corresponds exactly to ours, taking into account that $\phi=\pi/2-\varphi$ as well as their choice $\varphi_0=0$, $\theta_0=0$.

\subsection{Non-integer $\alpha$}
For non-integer $\alpha$, the periodicity that is used in arriving to (\ref{ft2d3}) from (\ref{ft2d2}) is lost. However, extending the domain of angular momentum variable to $\langle -\pi+i\infty, \pi + i\infty\rangle$, like shown in fig. 2, both periodicity and single-valuedness are recovered. In fact, there is an integral representation for Bessel's function of non-integer order (e.g. \cite{GR} pg 1402),

\begin{equation}\label{Bessel2}
J_\alpha(z)=\frac{1}{2\pi}\int_{-\pi+i\infty}^{\pi+i\infty} e^{-iz\sin\theta+i\alpha\theta}d\theta,
\end{equation} 
where the path of integration is shown in fig. 2., that allows one to conclude that the result from previous subsection also holds for a general (non-integer) $\alpha$, with the understanding that in that case the domain of the angle is to be extended like described. Separating the above integral into real and imaginary angle contributions,

\begin{equation}\label{Bessel2}
J_\alpha(z)=\frac{1}{2\pi}\left[\int_{-\pi}^{\pi} e^{-iz\sin\theta_R+i\alpha\theta_R}d\theta_R-2\sin(\alpha\pi)\int_0^\infty e^{z\sinh\theta_I-\alpha\theta_I} d\theta_I \right] ,
\end{equation}
where we introduced $\theta_R$ and $i\theta_I$ to distinguish the angular momentum variable on it's standard and extended domains, one immediately sees that the first contribution is the same as in the integer magnetic flux case, while the second contribution only adds in case of non-integer $\alpha$. Interpreting first term as incident, and second term as scattered wave, the scattering cross section proportionality with $\sin^2(\alpha\pi)$ is reproduced. Thus the imaginary part of the domain of the momentum angle, which does not have a classical analog, turns out responsible for the quantum scattering effect. 

We note that while the above argumentation is heuristic and rather indirect, the imaginary part of the domain appearing for non-integer $\alpha$  only when plane-waves are represented via Bessel functions, it is effectively $\mathbb{R_+\times R}$ geometry and therefore consistent with the geometry of it's inverse space, the configurational space.

\section{Experiment}
We have seen the appearance of Berry's phase on the wave-function of free flowing electrons. With respect to the original Berry's proposal, there is a possible practical benefit in that it is not necessary to localize electrons and move them around coil, but rather one can manipulate free-flowing electrons like in figure 3. 

The effect has already been proposed as an idea for quantum computing (check e.g. a pedagogical review in \cite{qc}). Interestingly, figure 3 may also represent a scheme of a prototype of a basic building block for the architecture of the quantum computer. Consider cases $\alpha=1/2, 1/3, 1/4$ etc as an example. When $\alpha$ is $1/2$ figure one is bimodal and corresponds to a classical computer  - the flows can be either in phase or anti-phase, depending on the evenness/oddness of the winding number. But when $\alpha$ is an inverse of an integer greater then two, the scheme in figure 3 is a multimodal logic. If it would be possible to incorporate stable current flows and coherent sources of free flowing electrons into a design with controllable back-reaction and dissipation effects, the scheme of figure 3 could perhaps indeed offer (one) basic solution for quantum computer. In addition, with the ability to variate the current flows in coils the logical structure of the machine becomes dynamic, allowing for locally switching between different logical modalities (like from quantum to classical computing) by locally manipulating current in a coil.   

\begin{figure}
\begin{center}
\begin{tikzpicture}[line cap=round,line join=round,>=triangle 45,x=1.0cm,y=1.0cm, scale=0.7]

\draw(-0.1,2.3) circle (2.09cm);
\draw [->] (1.1,3.2) -- (-1.4,3.2);
\draw [->] (1.1,2) -- (-1.4,2);
\draw [->] (1.1,2.6) -- (-1.4,2.6);
\draw [->] (1.1,1.4) -- (-1.4,1.4);
\draw [color=black, line width=1pt] (-0.1,2.3)-- ++(-1.5pt,-1.5pt) -- ++(3.0pt,3.0pt) ++(-3.0pt,0) -- ++(3.0pt,-3.0pt);

\draw(6.02,2.3) circle (2.09cm);
\draw [->] (7.2,3.2) -- (4.7,3.2);
\draw [->] (7.2,2) -- (4.7,2);
\draw [->] (7.1,1.6) -- (6.1,3.9);
\draw [->] (6,1.2) -- (5,3.5);
\draw [color=black, line width=1pt] (6.02,2.3)-- ++(-1.5pt,-1.5pt) -- ++(3.0pt,3.0pt) ++(-3.0pt,0) -- ++(3.0pt,-3.0pt);

\draw(-0.1,-3.4) circle (2.09cm);
\draw [->] (1.1,-2.5) -- (-1.4,-2.5);
\draw [->] (1.1,-3.7) -- (-1.4,-3.7);
\draw [->] (1,-2.1) -- (0,-4.4);
\draw [->] (-0.1,-1.7) -- (-1.1,-4);
\draw [color=black, line width=1pt] (-0.1,-3.4)-- ++(-1.5pt,-1.5pt) -- ++(3.0pt,3.0pt) ++(-3.0pt,0) -- ++(3.0pt,-3.0pt);

\draw(6.02,-3.4) circle (2.09cm);
\draw [->] (7.2,-2.5) -- (4.7,-2.5);
\draw [->] (7.2,-3.7) -- (4.7,-3.7);
\draw [->] (7.2,-3.1) -- (4.7,-3.1);
\draw [->] (7.2,-4.3) -- (4.7,-4.3);
\draw[line width=2.5pt] (4.3,-4.4)--(4.3,-2.4);
\draw [color=black, line width=1pt] (6.02,-3.4)-- ++(-1.5pt,-1.5pt) -- ++(3.0pt,3.0pt) ++(-3.0pt,0) -- ++(3.0pt,-3.0pt);

\draw [<-,shift={(6.08,2.38)}] plot[domain=4.46:5.41,variable=\t]({1*2.46*cos(\t r)+0*2.46*sin(\t r)},{0*2.46*cos(\t r)+1*2.46*sin(\t r)});
\draw [<-,shift={(0.04,-3.18)}] plot[domain=0.45:1.49,variable=\t]({1*2.22*cos(\t r)+0*2.22*sin(\t r)},{0*2.22*cos(\t r)+1*2.22*sin(\t r)});
\begin{scriptsize}
\fill [color=black] (1.1,3.2) circle (2pt);
\fill [color=black] (1.1,2) circle (2pt);
\fill [color=black] (7.2,3.2) circle (2pt);
\fill [color=black] (7.2,2) circle (2pt);
\fill [color=black] (1.1,-2.5) circle (2pt);
\fill [color=black] (1.1,-3.7) circle (2pt);
\fill [color=black] (7.2,-2.5) circle (2pt);
\fill [color=black] (7.2,-3.7) circle (2pt);
\fill [color=black] (1.1,2.6) ++(-2pt,0 pt) -- ++(2pt,2pt)--++(2pt,-2pt)--++(-2pt,-2pt)--++(-2pt,2pt);
\fill [color=black] (1.1,1.4) ++(-2pt,0 pt) -- ++(2pt,2pt)--++(2pt,-2pt)--++(-2pt,-2pt)--++(-2pt,2pt);
\fill [color=black] (7.1,1.6) ++(-2pt,0 pt) -- ++(2pt,2pt)--++(2pt,-2pt)--++(-2pt,-2pt)--++(-2pt,2pt);
\fill [color=black] (6,1.2) ++(-2pt,0 pt) -- ++(2pt,2pt)--++(2pt,-2pt)--++(-2pt,-2pt)--++(-2pt,2pt);
\fill [color=black] (1,-2.1) ++(-2pt,0 pt) -- ++(2pt,2pt)--++(2pt,-2pt)--++(-2pt,-2pt)--++(-2pt,2pt);
\fill [color=black] (-0.1,-1.7) ++(-2pt,0 pt) -- ++(2pt,2pt)--++(2pt,-2pt)--++(-2pt,-2pt)--++(-2pt,2pt);
\fill [color=black] (7.2,-3.1)++(-2pt,0 pt) -- ++(2pt,2pt)--++(2pt,-2pt)--++(-2pt,-2pt)--++(-2pt,2pt);
\fill [color=black] (7.2,-4.3) ++(-2pt,0 pt) -- ++(2pt,2pt)--++(2pt,-2pt)--++(-2pt,-2pt)--++(-2pt,2pt);
\end{scriptsize}
\end{tikzpicture}
\end{center}
\caption{A schematic depiction of an experiment aimed at measuring Berry's phase. Two sets of coherent sources of electrons, depicted by $\bullet$ and $\blacklozenge$, emit electrons in a common direction, on a point magnetic flux, depicted with $\times$, like in the first picture. Then one set of sources is rotated around for a full circle, while the other remains static, keeping them at all times coherent. Finally, the two sets of flowing electrons are superimposed on an interference screen, as in the last picture. Due to translational symmetry, the two flows need not be in the same plane, and can be separated by an insulator until the final stage when the beams are superimposed.} \label{fig:M3}
\end{figure}
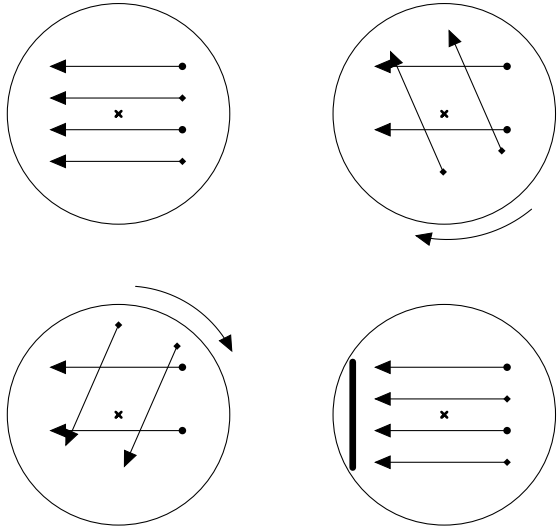

\section{Conclusion and outlook}
Formulation of the problem on a non-trivial background geometry makes it possible to obtain Berry's phase by simple integration, as shown in this paper. It also allowed us to obtain a particular solution for the scattering on multiple coils. In the latter case, we weren't able to define the background geometry, which will be the topic of further investigations.

Analyzing the single-coil solution in the momentum representation, we found that the phase doesn't depend on the realization. We also found that the domain of the momentum angular component gets extended, in correspondence with the same extension in the configuration space.

Finally, from practical point of view, this formulation treats motion of free electrons on the same footing as global motions of the system. The wave-function of the electron only records the total motion it made on the configuration space, regardless of the cause. We offer a scheme for experimental verification of this fact.

\section*{Acknowledgment}
I'm grateful to S. Mignemi and D. Vale for reading the manuscript.


\begin{thebibliography}{9}

\bibitem{AB}
  Y. Aharonov, D. Bohm, Phys. Rev. \textbf{115} 3 (1959)
\bibitem{ber} M. V. Berry, Proc. R. Soc. Lond. A \textbf{392} 45-57 (1984)
\bibitem{sim} B. Simon, Phys. Rev. Lett. \textbf{51} 2167 (1983)
\bibitem{wilczek} F. Wilczek, Phys. Rev. Lett. \textbf{48} 17 (1982); Phys. Rev. Lett. \textbf{49} 14 (1982)
\bibitem{jack}  R. Jackiw, Int. J. Mod. Phys. A. \textbf{3} 285 (1988)
\bibitem{BF} M. Born, V. Fock, Zs. Phys \textbf{51} 165 (1928)
\bibitem{dirac} P. A. M. Dirac, Proc. R. Soc. Lond. A \textbf{133} 60-72 (1931)
\bibitem{revija} D. Xiao, M.-C. Chang, Q. Niu, Rev. Mod. Phys. \textbf{82} 3 (2010)
\bibitem{knjiga} A. Bohm, A. Mostafazadeh, H. Koizumi, Q. Niu, J. Zwanziger, \textit{The Geometric Phase in Quantum Systems}, Springer (2007)
\bibitem{qc} G. K. Brennen, J. K. Pachos, Proc. R. Soc. A \textbf{464} 1-27 (2008)
\bibitem{nambu} Y. Nambu, Nucl. Phys. \textbf{B579} 590-616 (2000)
\bibitem{sto} P. \v S\v tovi\v cek, Phys. Lett. A \textbf{142} 1 (1989)
\bibitem{ganga} S. A. H. Gangaraj, M. G. Silverinha, G. W. Hanson, JMMTP \textbf{2} (2017)
\bibitem{drago} D. Dragoman, S. Bogdan, Opt. Comm. \textbf{281} 9 (2008)
\bibitem{AA} Y. Aharonov, J. Anandan, Phys. Rev. Lett. \textbf{58} 1593-1596 (1986)
\bibitem{GR} I. S. Gradshteyn, I. M. Ryzhik, \textit{Table of Integrals, Series, and Products}, AP (1996)

\end{thebibliography}
\end{document}